%% file: main.tex
\documentclass[letterpaper, 10 pt, conference]{ieeeconf}
\IEEEoverridecommandlockouts
\let\labelindent\relax
\usepackage{enumitem}
\usepackage{balance}
\usepackage{subcaption}
\usepackage[font={small}]{caption}
\usepackage{array}
\usepackage{textcomp}
\usepackage{mathtools, nccmath}
\usepackage{graphicx}
\usepackage{amsfonts}
\usepackage{amsmath}
\usepackage{amssymb}
\usepackage{algorithm}
\usepackage{algpseudocode}
\usepackage{hyperref}
\usepackage{tikz}
\usepackage{arydshln}
\usepackage{multirow}
\usepackage{bm}
\usepackage{epstopdf}
\usepackage{cite}
\usepackage{siunitx}
\usepackage{bbm}

\DeclareMathOperator*{\argmin}{argmin} % no space, limits underneath in displays
\newtheorem{remark}{Remark}

\begin{document}
\title{\LARGE \bf i2LQR: Iterative LQR for Iterative Tasks in Dynamic Environments}

\author{Yifan Zeng$^{*1}$, Suiyi He$^{*2}$, Han Hoang Nguyen$^3$, Yihan Li$^4$, Zhongyu Li$^3$, Koushil Sreenath$^3$ and Jun Zeng$^3$
\thanks{$^*$Authors have contributed equally.}
\thanks{Implementation code is released on \url{https://github.com/HybridRobotics/ilqr-iterative-tasks}.}
\thanks{$^1$Author is with Shanghai Jiao Tong University, Shanghai, China. {\tt\small blakezyf1107@sjtu.edu.cn}}
\thanks{$^2$Author is with University of Minnesota-Twin Cities, MN 55455, USA. {\tt\small he000231@umn.edu}}
\thanks{$^3$Authors are with University of California, Berkeley. {\tt\small \{hanhn, zhongyu\_li,koushils,zengjunsjtu\}@berkeley.edu}}
\thanks{$^4$Author is with Xi'an Jiaotong University, Xi'an, China. {\tt\small 1325140363@stu.xjtu.edu.cn}}
}

\maketitle
\input{sections/abstract}
\IEEEpeerreviewmaketitle
\input{sections/introduction}
\input{sections/background}
\input{sections/algorithm}

\input{sections/results}
\input{sections/conclusion}
\bibliographystyle{IEEEtran}
\balance
\bibliography{references}{}
\end{document}

%% file: sections/abstract.tex
\begin{abstract}
This work introduces a novel control strategy called Iterative Linear Quadratic Regulator for Iterative Tasks (i2LQR), which aims to improve closed-loop performance with local trajectory optimization for iterative tasks in a dynamic environment.
The proposed algorithm is reference-free and utilizes historical data from previous iterations to enhance the performance of the autonomous system.
Unlike existing algorithms, the i2LQR computes the optimal solution in an iterative manner at each timestamp, rendering it well-suited for iterative tasks with changing constraints at different iterations.
To evaluate the performance of the proposed algorithm, we conduct numerical simulations for an iterative task aimed at minimizing completion time.
The results show that i2LQR achieves an optimized performance with respect to learning-based MPC (LMPC) as the benchmark in static environments, and outperforms LMPC in dynamic environments with both static and dynamics obstacles. 
\end{abstract}

%% file: sections/introduction.tex
\section{Introduction}\label{sec:introduction}
\subsection{Motivation}
One important objective of control algorithms is to optimize the performance of autonomous systems.
This can be usually formulated as minimizing completion time \cite{jain2020computing} or energy consumption \cite{wu2022model} during task execution. 
The performance-optimal controller can be applied to iterative tasks, where the autonomous system must do the same task repeatedly.
However, the surrounding environment around the autonomous system may be dynamic. This means that constraints of the system are not always the same along the process, such as constraints that are changed in some given iterations, e.g. new obstacles appear after a particular iteration or moving obstacles in a single iteration.
The existing algorithms cannot handle these changes effectively. This motivates us to propose a control strategy that could handle additional constraints in different iterations while improving the system's closed-loop performance with local trajectory optimization.

\subsection{Related Work}
Researchers have applied different methods to optimize performance. 
Model-based approaches usually leverage a high-level planner to generate the optimized trajectory and a low-level controller is deployed to track the planned trajectory \cite{kapania2016sequential,nagy2019sequential, heilmeier2019minimum, palleschi2021fast}.
However, calculating the best possible trajectory can be time consuming, and the low-level controller may not track the trajectory perfectly due to discretization in the planning problem. 
Furthermore, when the planned trajectory encounters conflicts with newly appeared constraints, local re-planning is required to deal with these additional constraints~\cite{gao2020teach}.
This may result in the loss of optimality for the trajectory in local trajectory optimization.

Several attempts are made to address the problem through data-driven based model-free approaches, where the system's historical data is used to train an end-to-end control policy directly.
For instance, in \cite{jain2020computing,fuchs2021super}, the learning-based control policy shows its capability to push an autonomous racing car to its dynamics limit.
In \cite{song2021autonomous, penicka2022learning}, quadrotors are shown to fly with aggressive maneuvers in an autonomous drone racing competition.
Nevertheless, these methods still have limitations.
Learning-based methods are data hungry and require significant time to get the policy, which means that such methods may not be suitable for some real-time applications.
Moreover, all the aforementioned policies operate in a static environment.
However, in practice, the control policy is required to be functional in a dynamic environment.
More importantly, since the performance of the trained policy is highly related to the used training data set, the learned policy may not guarantee optimal performance.

Recently, reference-free, model-based methods are proposed to provide optimized performance for iterative tasks. 
In \cite{kabzan2019learning}, a model predictive contouring controller (MPCC) with dynamics learning is used for autonomous racing cars. 
The optimization algorithm called learning-based MPC (LMPC) is introduced in \cite{rosolia2017learning, rosolia2021minimum}, where the system's historical data is used to formulate the local MPC optimization problem.
This allows the system to improve its performance over each iteration and the system is proved theoretically to achieve closed-loop optimal performance \cite{rosolia2017learning}.
The proposed algorithm is implemented on autonomous vehicles \cite{rosolia2019learning}, aerial robots \cite{li2022learning} and robotic arms \cite{thananjeyan2021abc}.
However, these methods still have shortcomings.
Although the local objective function in \cite{kabzan2019learning} considers optimal performance along the prediction horizon, the result is not closed-loop optimal without global planning.
The LMPC strategy in \cite{rosolia2017learning, rosolia2021minimum} must work in a static environment, which means that the scenario should be exactly same for every iteration.
This is due to the limitation of the optimization setup, where local MPC's feasibility strictly depends on the reachability of historical states from previous iterations.
If new obstacles are introduced, these historical states could be infeasible, rendering the MPC problem infeasible.
One possible solution to this problem is local-replanning \cite{he2022autonomous},
but the performance may be limited due to the nonsmooth switching in the high-level planner.

To ensure that the autonomous system can smoothly adapt to newly introduced constraints, the optimal trajectory should be computed in an iterative manner.
Therefore, we want to use a method that could handle this without resulting in infeasibility.
The iterative linear quadratic regulator (iLQR) shows potential in solving this problem.
iLQR is an extension of LQR control, where the optimization is solved iteratively and linearization of cost function and system dynamics is conducted in each iteration.
In \cite{chen2017constrained, hu2022active}, the iLQR computes the open-loop predicted trajectory for control problems in an iterative manner.
This motivates us to investigate the above challenging problem using an iLQR-based algorithm. 

\subsection{Contribution}
The contributions of this paper are as follows:
\begin{itemize}
    \item We propose a novel optimal control strategy called Iterative LQR for Iterative Tasks (i2LQR), which improves closed-loop performance with local trajectory optimization for a general dynamic system for iterative tasks in dynamic environments.
    %\item We show how to use historical data from previous iterations to build the local optimization problem of i2LQR at each time step and how to solve this optimization problem in an iterative manner.
    \item We demonstrate the utilization of historical data from preceding iterations to formulate the local optimization problem within each time step of i2LQR. Additionally, we outline an iterative approach to efficiently solve this optimization problem.
    \item Through numerical simulation, our proposed control strategy is shown to achieve the same optimal performance as the LMPC algorithm for iterative tasks in static environments and outperform the LMPC algorithm for iterative tasks in dynamic environments. 
\end{itemize}

%% file: sections/background.tex
\section{Problem Setup}
\label{sec:setup}
In this section, we show the problem setup of iterative tasks.
\begin{table}[t!]
    \setlength{\abovecaptionskip}{0.5cm}
    \setlength{\belowcaptionskip}{-0.5cm}
    \centering
    \scriptsize
    \caption{Symbol Notations}
    \vskip-10pt
    \begin{tabular}{c | c}\hline
        \multicolumn{2}{c}{\textbf{Symbols for iterative tasks}}\\\hline
        Symbol & Description\\ \hline
        $\mathbf{x}_t^i$  & System state at time step $t$ of iteration $i$\\\hline
        $\mathbf{u}_t^i$ & System input at time step $t$ of iteration $i$\\\hline
        $\mathbf{x}_0$ & Initial state for all iterations\\ \hline
        $\mathbf{x}_{\text{target}}$ & Target state for all iterations\\\hline
        $h(\mathbf{x}^i_t)$ & Cost-to-go associated with system state $\mathbf{x}^j_t$ \\ \hline
        $\mathcal{X}^{i}$ & Set of historical states for iteration $i$\\ \hline
        $\mathcal{H}$ & Set of all historical states from previous iterations \\\hline
        $\mathcal{C}_{t}^i$ & Constraints on the system at time step $t$ in iteration $i$ \\ \hline
        $\epsilon$ & A small positive number \\ \hline
        \multicolumn{2}{c}{\textbf{Symbols for i2LQR}}\\\hline
        Symbol & Description\\ \hline
        $\bar{\mathbf{x}}_{r}$ & Guided state for $r$-th target terminal set \\\hline
        $\mathcal{Z}_{r}$ & $r$-th target terminal set\\\hline
        $\mathbf{z}_{r}(j)$ & $j$-th state from $r$-th target terminal set \\\hline
        $N$ & Prediction horizon for the optimization problem \\\hline       
        ${\mathbf{x}}_{r}^m(j)$& Open-loop states of $m$-th iteration of iLQR for $\mathbf{z}_{r}(j)$\\\hline
        ${\mathbf{u}}_{r}^m(j)$
        & Open-loop inputs of $m$-th iteration of iLQR for $\mathbf{z}_{r}(j)$\\\hline
        ${\mathbf{x}}_{r}^*(j)$
        & Optimized open-loop states from iLQR for $\mathbf{z}_{r}(j)$\\\hline
        ${\mathbf{u}}_{r}^*(j)$
        & Optimized open-loop inputs from iLQR for $\mathbf{z}_{r}(j)$\\\hline
        $\mathbf{x}^*_r(j_r^*)$ & Best open-loop states for $r$-th optimization cycle\\\hline
        $\mathbf{u}^*_r(x_r^*)$ & Best open-loop inputs for $r$-th optimization cycle\\\hline  
        $J_{\mathbf{z}}(\bar{\mathbf{x}}_{r})$ &  Cost for nearest point selection on $\bar{\mathbf{x}}_{r}$\\\hline
        $J_{l}(\mathbf{x}_t^i,\mathbf{z}_{r}(j))$ &  Cost for local iLQR Optimization\\\hline
        $m_\text{max}$ & Maximum iteration for iterative optimization \\\hline
        $r_\text{max}$ & Maximum iteration for target terminal set \\\hline
    \end{tabular}
    \label{tab:setup-parameter}
    \vskip-10pt
\end{table}
For an iterative task at time step $t$ of iteration $i$, an autonomous system with dynamics $\mathbf{x}^i_{t+1} = f(\mathbf{x}^i_{t}, \mathbf{u}^i_{t})$ performs the task repeatedly until completion.
In each iteration, the system starts from the same initial state $\mathbf{x}_0$ and ends at the same target state $\mathbf{x}_\text{target}$.
The system's historical data (e.g. state and completion time) is saved in a data set $\mathcal{H}$.
The system's controller computes the optimal input based on this data set $\mathcal{H}$, state $\mathbf{x}^i_t$ and constraint $\mathcal{C}_t^i$ at current time step.
The constraint $\mathcal{C}_t^i$ includes the constraints on both system states and inputs.
If the system works in a static environment \cite{rosolia2021minimum, rosolia2019learning}, the constraint $\mathcal{C}_t^i$ will be same for the same time step $t$ of different iterations (i.e., $\mathcal{C}_t^i=\mathcal{C}_t^{i+1}$).
If the system works in a dynamic environment (as in this work), the constraint $\mathcal{C}_t^i$ will change along the entire process (i.e., $\mathcal{C}_t^i\neq\mathcal{C}_t^{i+1}$ or $\mathcal{C}_{t+1}^i\neq\mathcal{C}_t^{i}$).
The cost-to-go $h(\textbf{x}_t^i)$ is the time to finish the iteration $i$ from the point $\textbf{x}_t^i$ to $\mathbf{x}_{\text{target}}$, which means that the algorithm aims to minimize the time to finish each iteration.
This describes the cost to finish the corresponding iteration from that point to the target state $\mathbf{x}_{\text{target}}$.
The general form of this problem is shown in Alg. \ref{alg:iterative-task}. 

In this work, we focus on designing a performance optimal controller, which could minimize the cost $h(\mathbf{x}^i_0)$ of the iterative task. Details about this algorithm will be presented in Sec.~\ref{sec:algorithm}.
Parameters used in this work are listed in TABLE \ref{tab:setup-parameter} along with their notations.

\begin{figure*}[h]
    \centering
    \begin{subfigure}[t]{1.0\linewidth}
        \centering
        \includegraphics[width=0.8\linewidth]{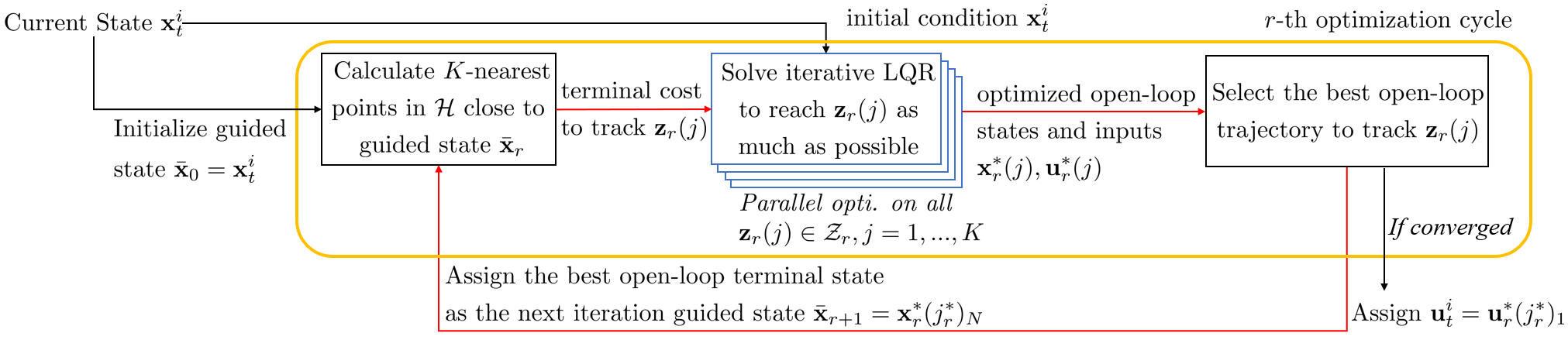}
        \caption{Proposed i2LQR. The optimization problem is resolved iteratively through several optimization cycles colored in the yellow block, and multiple iLQR problems are solved in parallel at each optimization cycle colored in blue.}
        \label{fig:diagram-i2LQR}
    \end{subfigure}
    \begin{subfigure}[t]{1.0\linewidth}
        \centering
        \includegraphics[width=0.8\linewidth]{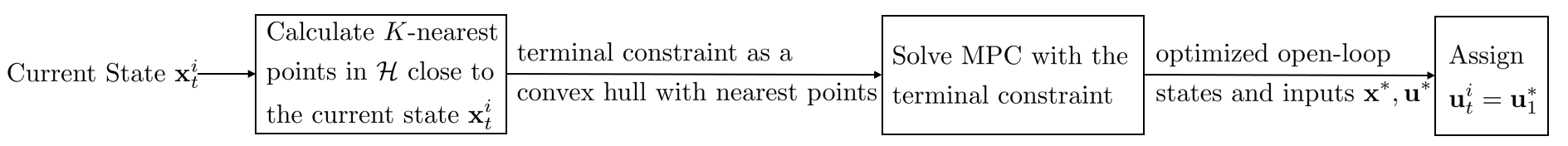}
        \caption{LMPC~\cite{rosolia2017learning,rosolia2021minimum}. The key difference between LMPC and our proposed i2LQR is that performance optimal points are regarded as terminal constraints instead of terminal costs updated iteratively.}
        \label{fig:diagram-LMPC}
    \end{subfigure}
   \caption{Illustration of i2LQR and existing LMPC algorithms}
    \label{fig:strategy}
\end{figure*}

%% file: sections/algorithm.tex
\section{Algorithm}
\label{sec:algorithm}

After introducing the problem setup for iterative tasks, the design of the proposed Iterative LQR for Iterative Tasks (i2LQR) in a dynamic environment will be presented in this section.
The general idea of the algorithm will be introduced in Sec.~\ref{sec:algorithm_overview}.
Details about the i2LQR will be shown in the following subsections.

\begin{algorithm}[b!]
\caption{Iterative Tasks}  
\label{alg:iterative-task}  
\begin{algorithmic}[1]
\State $\mathcal{H} \leftarrow \emptyset$
\Repeat
\State Iteration $i$ begins, $t \leftarrow 0$, $\mathbf{x}^i_0 \leftarrow \mathbf{x}_0$, $\mathcal{X}^i\leftarrow\mathbf{x}^i_0$
\While{$||\mathbf{x}^i_t - \mathbf{x}_{\text{target}} ||_2 \geq \epsilon$}
\State $\mathbf{u}^i_t \leftarrow \text{Controller}(\mathbf{x}^i_t, \mathcal{C}^i_t, \mathcal{H}$) \label{alg:iterative-task-controller}
\State $\mathbf{x}^i_{t+1}\leftarrow f(\mathbf{x}^i_t, \mathbf{u}^i_t)$
\State $t \leftarrow t+1, \mathcal{X}^i \leftarrow \mathcal{X}^i \cup \mathbf{x}^i_t$
\EndWhile
\State $\mathcal{H} \leftarrow \mathcal{H} \cup \mathcal{X}^i, i \leftarrow i+1$
\Until{Task is finished}
\end{algorithmic}  
\end{algorithm}

\subsection{Structure of i2LQR} \label{sec:algorithm_overview}
Consider the problem of achieving the system's optimal performance for iterative tasks in a dynamic environment in the form of Alg. \ref{alg:iterative-task}.
The proposed control strategy i2LQR computes the optimal input using historical data from previous iterations.
This includes states that the system has visited in previous iterations and the cost-to-go $h(\mathbf{x}^i_t)$ associated with each historical state.
In the first iteration, any open-loop controller could be used to generate a feasible trajectory as the initial history data.
Then, the proposed i2LQR algorithm is deployed to calculate the system's optimal input at each time step.
Fig.~\ref{fig:diagram-i2LQR} shows the structure of the proposed algorithm.
As a comparison, the structure of the LMPC algorithm is also presented in Fig.~\ref{fig:diagram-LMPC}.

\begin{figure}
    \setlength{\abovecaptionskip}{0.35cm}
    \setlength{\belowcaptionskip}{-0.5cm}
    \centering
    \includegraphics[width=0.85\linewidth]{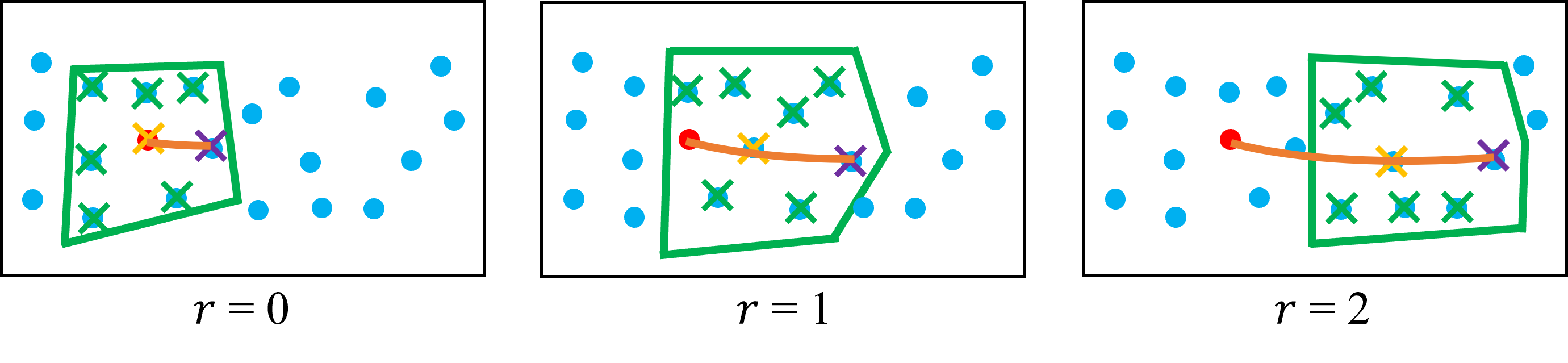}
    \caption{An illustration of nearest points selection in an iterative manner for different optimization cycles. 
    System state at the current time step is marked in red, while historical states are marked in blue. 
    States on the right come with a smaller cost-to-go.  
    Points with crosses in the green block are the target terminal set $\mathcal{Z}_r$, consisting of the selected $K$ nearest points, in which $\mathbf{z}_{r}(j)( j=1,2,...,K)$ represents a single terminal state(a single point with cross).
    Specifically, yellow is the guided state, and purple is the state associated with the best open-loop trajectory. 
    The orange line is the best open-loop trajectory.}
    \label{fig:iterative_point_selection}
\end{figure}

Different from LMPC algorithm, the proposed i2LQR controller consists of several optimization cycles (yellow block in Fig.~\ref{fig:diagram-i2LQR}).
For the $r$-th optimization cycle, we define a guided state $\bar{\mathbf{x}}_{r}$, which will be the initial guess of the open-loop terminal state in the $r$-th optimization cycle and we select $K$ nearest points with respect to it (see Fig.~\ref{fig:iterative_point_selection}).
For the first cycle, the state at the current time step $\mathbf{x}^i_t$ will be used as the guided state, while the best open-loop predicted terminal state ${\mathbf{x}}_{r-1}^*(j_{r-1}^*)_N$ from the last cycle will be used as the guided state for $r$-th cycle.
Then $K$-nearest points to $\bar{\mathbf{x}}_{r}$ from the historical states set $\mathcal{H}$ will build the target terminal set $\mathcal{Z}_r$, which consists of the target terminal state $\mathbf{z}_r(j)$ of the $j$-th iLQR optimization.
To reduce the computational time of the proposed algorithm, these iLQR optimizations will be solved through parallel computing, colored in blue in Fig.~\ref{fig:diagram-i2LQR}.
Then the best open-loop predicted solution for the $r$-th optimization cycle will be selected.
The algorithm will continue doing optimization until either the set $\mathcal{Z}_r$ remains unchanged or the maximum cycle number $r_{\text{max}}$ is reached.
As a result, the algorithm will select the iLQR's optimal target terminal state $\mathbf{z}_r(j_r^*)$ from historical data in an iterative manner.
Details about this algorithm will be illustrated in the following subsections.

\subsection{Nearest Points Selection}\label{sec:algorithm-nearest-point}
To build the target terminal set $\mathcal{Z}_r$, the following criteria will be used to select the $K$-nearest points:
\begin{subequations}\label{eq:nearest_points}
\begin{align}
    J_\mathbf{z}(\bar{\mathbf{x}}_r)  = \min\limits_{\mathbf{z}_r(j)} 
    &{\sum\limits_{j=1}\limits^{K}||{\mathbf{z}_{r}(j) - \bar{\mathbf{x}}_r}}||^{2}_{D_0} \\
    \text{s.t.}\quad z_{j} \neq z_{l}, &~\forall{j\neq l}\\
     z_{j} \in \mathcal{H},&~j= 1,...,K 
\end{align}
\end{subequations}
where $j$ refers to the index of points in the $r$-th optimization cycle;
$D_0$ is a diagonal matrix that contains weighting factors for state variables.

\begin{remark}
The number of maximum iterations $r_{\text{max}}$ and the number of selected nearest points $K$ are the hyperparameters of the proposed algorithm. 
A larger value of $r_{\text{max}}$ or $K$ will make the system converge to the optimal performance more quickly.
However, this will also increase the computational burden at each time step.
\end{remark}

\subsection{Local iLQR Optimization} \label{sec:algorithm_optimization_design}
The following constrained finite-time optimal control problem will be solved through iLQR for each $\mathbf{z}_r(j)$:
\begin{subequations}
\label{eq:i2lqr}
\begin{align}
J_l(\mathbf{x}_t^i, \mathbf{z}_r(j)) = \min\limits_{{\mathbf{u}}^*_{r}(j)} p(\mathbf{x}^*_r(j)_{1+N}, \mathbf{z}_r(j)&) \label{eq:i2lqr-cost}\\
\text{s.t.} \quad 
    \mathbf{x}^*_{r}(j)_{k+1} = f(\mathbf{x}^*_{r}(j)_{k}, \mathbf{u}^*_{r}(j)_{k}), k& = 1,...,N\label{eq:i2lqr-dynamics} \\
    \mathbf{x}^*_{r}(j)_{k+1} ,\mathbf{u}^*_{r}(j)_{k} \in \mathcal{C}^i_{t+k|t}, k& = 1,...,N \label{eq:i2lqr-constraint}\\
    \mathbf{x}^*_{r}(j)_{1} = \mathbf{x}^i_t,~~~~~~~~~~~&  \label{eq:i2lqr-initial-condition}
\end{align}
\end{subequations}
where \eqref{eq:i2lqr-cost} is the optimization problem's objective function;
\eqref{eq:i2lqr-dynamics} represents the system dynamics;
\eqref{eq:i2lqr-constraint} shows the constraints of the system along the prediction horizon;
\eqref{eq:i2lqr-initial-condition} is the initial constraint.
Specifically, the terminal cost introduces the difference between the open-loop predicted terminal state and the target terminal state in the quadratic form:
\begin{equation}\label{eq:i2lqr-terminal-cost}
\begin{aligned}
    p(\mathbf{x}^*_r(j)_{1+N}, \mathbf{z}_r(j))&=\\(\mathbf{x}^*_r(j)_{1+N}& - \mathbf{z}_r(j))^{T}P(\mathbf{x}^*_r(j)_{1+N}- \mathbf{z}_r(j)),
\end{aligned}
\end{equation}
where $P$ is a diagonal matrix consisting of weighting factors.

\begin{algorithm}[t!]
\caption{iLQR}  
\label{alg:iLQR}  
\begin{algorithmic}[1]
\State $\mathbf{u}_{r}^0(j) \leftarrow \mathbf{0}$ 
\Repeat
\State Iteration $m$ begins
\State $\mathbf{x}_{r}^m(j)_\leftarrow g(\mathbf{x}_t^i,\mathbf{u}_{r}^m(j))$ \label{alg:forward_dynamics}
\State Linearize $f(\cdot)$ and $J_s(\cdot)$ around $\mathbf{x}_{r}^m(j)$ and $\mathbf{u}_{r}^m(j)$\label{alg:linearization}
\State $\delta^*(\mathbf{u}_{r}^m(j))\leftarrow \text{LQR}(\delta(\mathbf{x}_{r}^m(j)), \delta(\mathbf{u}_{r}^m(j)))$
\State $\mathbf{u}_{r}^{m+1}(j)\leftarrow\mathbf{u}_{r}^m(j)+\delta^*(\mathbf{u}_{r}^m(j)), m \leftarrow m+1$
\Until{Reach $m_\text{max}$ OR $J_s(\cdot)$ has converged}
\end{algorithmic}  
\end{algorithm}

Alg.~\ref{alg:iLQR} shows how to solve the above optimization problem through iLQR.
System constraints along the prediction horizon will be converted to part of the new cost function $J_s(\cdot)$ through the exponential function as done in \cite{chen2017constrained}.
Firstly, the algorithm starts with an initial input sequence, such as zero control inputs in this work.
Then, $g(\cdot)$ will calculate the open-loop states based on open-loop inputs and initial state through system dynamics \eqref{eq:i2lqr-dynamics} during the forward pass at line \ref{alg:forward_dynamics}.
It will be linearized along with the cost function $J_s(\cdot)$ around $\mathbf{x}_{r}^m(j)$ and $\mathbf{u}_{r}^m(j)$.
The optimal solution $\delta^*(\mathbf{u}_{r}^m(j))$ could be obtained efficiently and is used to generate the input sequence $\mathbf{u}_{r}^{m+1}(j)$ for the next iteration.
The algorithm will do this computation repeatedly until the cost $J_s(\cdot)$ has converged or the maximum iteration $m_{\text{max}}$ is reached.

\subsection{Best Open-Loop Solution Selection}\label{sec:algorithm-optimal-selection}
In each optimization cycle, the best open-loop solution will be selected among $K$ solutions. 
The following local cost will be used in this process:
\begin{equation}\label{eq:terminal-state-local-cost}
    j^*_r{=}\argmin_{j = 1, ..., K} w_h h(\mathbf{z}_r(j_r)) + w_d ||\mathbf{x}^*_r(j_r)_N- \mathbf{z}_r(j_r)||^2_{D_1},
\end{equation}
where $w_h$, $w_d$ are weighting factors;
$h(.)$ is the cost-to-go associated with the state $\mathbf{z}_r(j_r)$;
$||\mathbf{x}^*_r(j_r)_N- \mathbf{z}_r(j_r)||^2_{D_1}$ describes the penalty for the state difference between $\mathbf{x}^*_r(j_r)_N$ and $\mathbf{z}_r(j_r)$ with a diagonal weighting matrix $D_1$.
Since cost-to-go function $h(.)$ is considered in~\eqref{eq:terminal-state-local-cost}, this allows the iterative optimization to find the best terminal state in the outer loop of i2LQR, shown in red connected lines in Fig.~\ref{fig:diagram-i2LQR}.

%% file: sections/results.tex
\begin{figure}
    \setlength{\abovecaptionskip}{0.3cm}
    \setlength{\belowcaptionskip}{-0.5cm}
    \centering
    \begin{subfigure}[t]{1\linewidth}
        \setlength{\abovecaptionskip}{0cm}
        \setlength{\belowcaptionskip}{0.1cm}
        \centering
        \includegraphics[width=0.94\linewidth]{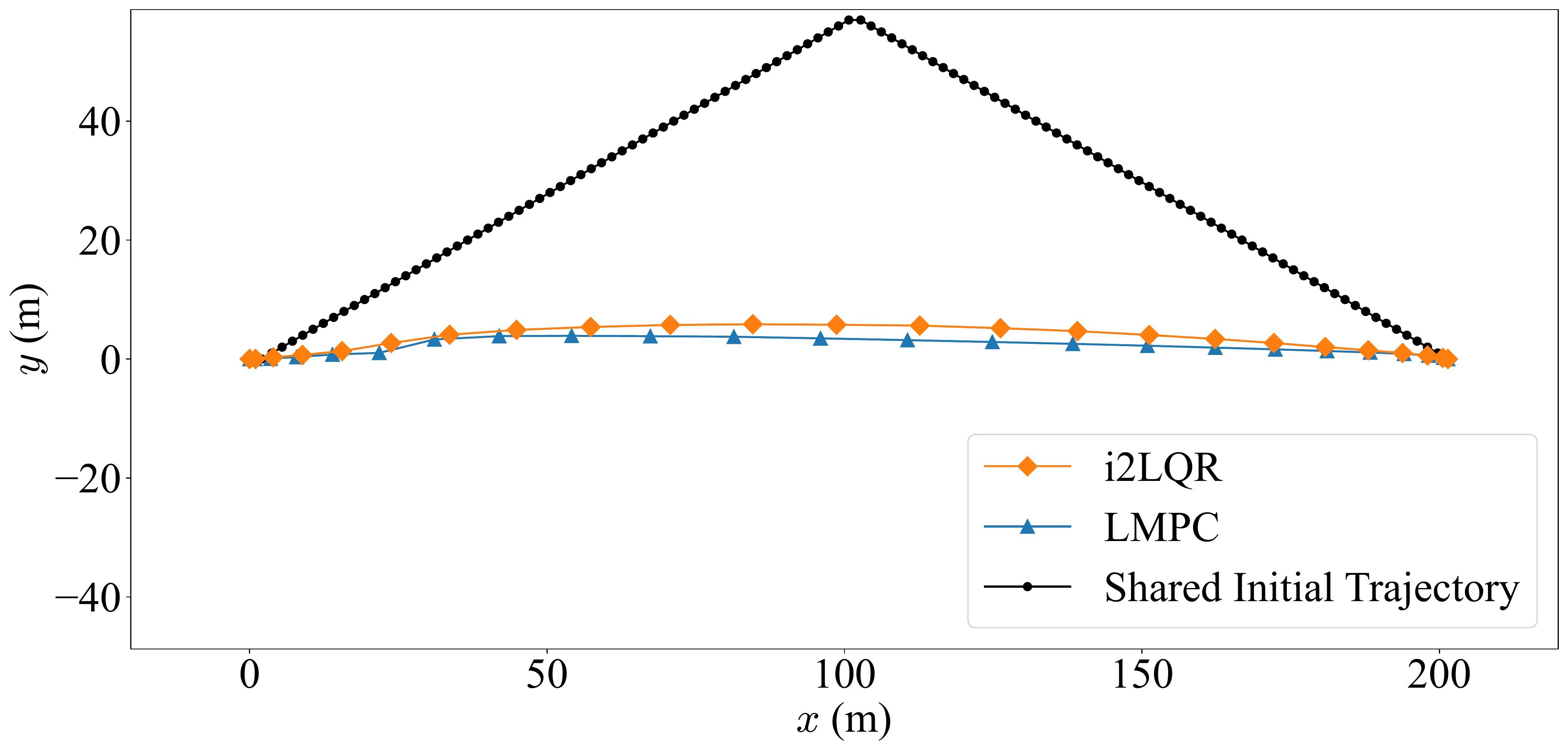}
        \caption{Shared initial trajectory, and trajectories in iteration 10 for i2LQR and LMPC.}
        \label{fig:no_obstacle_trajectory}
    \end{subfigure}
    \begin{subfigure}[t]{1\linewidth}
        \setlength{\abovecaptionskip}{0cm}
        \setlength{\belowcaptionskip}{0.1cm}
        \centering
        \includegraphics[width=0.94\linewidth]{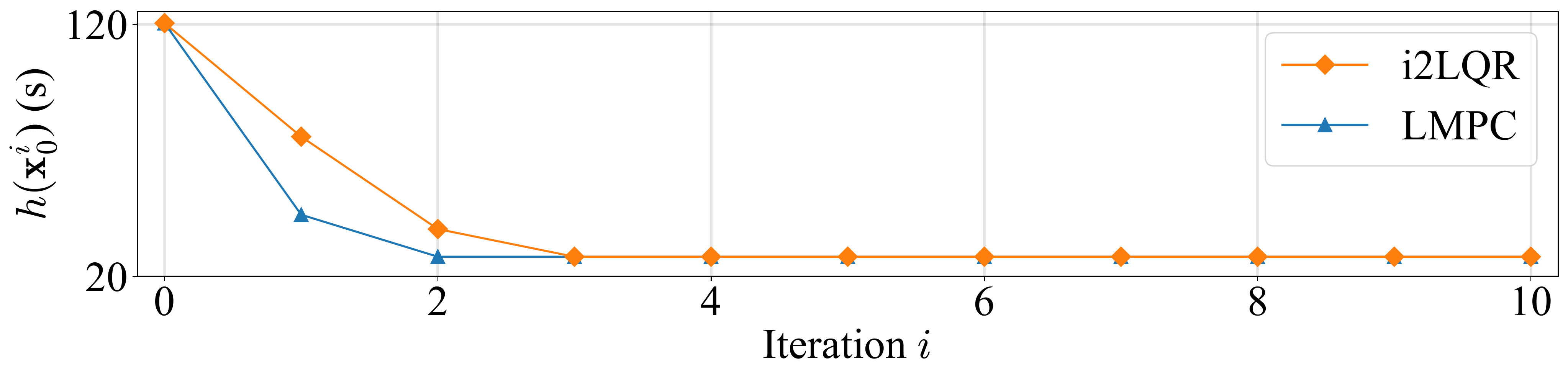}
        \caption{Time to finish the iteration for i2LQR and LMPC.}
        \label{fig:no_obstacle_time}
    \end{subfigure}
    \begin{subfigure}[t]{1\linewidth}
        \setlength{\abovecaptionskip}{0cm}
        \setlength{\belowcaptionskip}{0.1cm}
        \centering
        \includegraphics[width=0.94\linewidth]{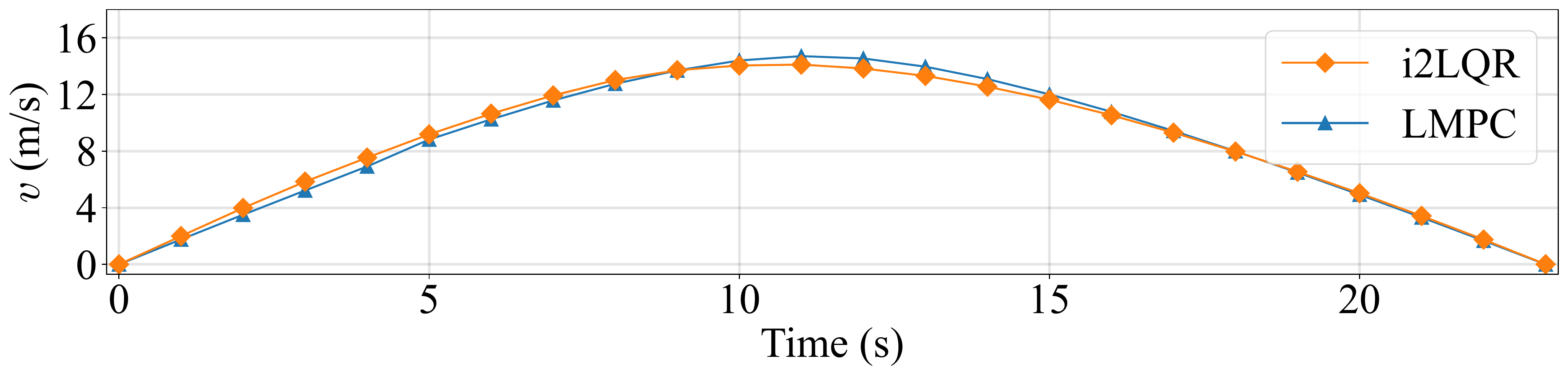}
        \caption{System speed in iteration 10 for i2LQR and LMPC.}
        \label{fig:no_obstacle_spd}
    \end{subfigure}
    \begin{subfigure}[t]{1\linewidth}
        \setlength{\abovecaptionskip}{0cm}
        \setlength{\belowcaptionskip}{0.1cm}
        \centering
        \includegraphics[width=0.94\linewidth]{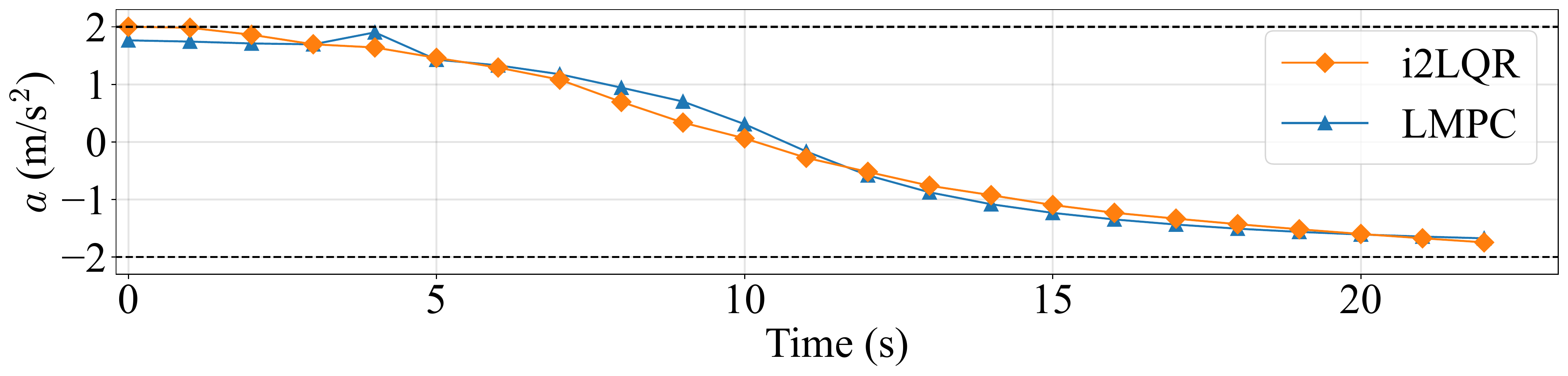}
        \caption{System acceleration in iteration 10 for i2LQR and LMPC.}
        \label{fig:no_obstacle_acc}
    \end{subfigure}
   \caption{Simulation with no obstacle. Both algorithms could reach the system's optimal performance.}
    \label{fig:sim-1}
\end{figure}

\section{Results}
\label{Sec:Results}
Having presented the framework that uses i2LQR for iterative tasks in dynamic environments in the previous sections, we now show the performance of the proposed algorithm. 
In Sec. \ref{Sec:Results-Setup}, the simulation setup is introduced.
Then, the performance of the proposed controller is compared with state-of-the-art learning-based MPC algorithm for iterative tasks in both static environments and dynamic environments.

\subsection{Simulation Setup}\label{Sec:Results-Setup}
A nonlinear kinematic bicycle model with input constraints as in \cite{chen2017constrained} is used to evaluate the proposed algorithm. The kinematic bicycle model has states and inputs at time step $t$ given by $\textbf{x}_t=[x_t, y_t, v_t, \theta_t]^T, \textbf{u}_t=[a_t, \delta_t]^T$,
where $x_t$ and $y_t$ describe the system's position;
$v_t$ and $\theta_t$ show the system's speed and heading angle;
$a_t$ is the acceleration;
$\delta_t$ represents the steering angle.
The sampling time $\Delta t$ is set to 1 second, which is consistent with the open-source code of~\cite{rosolia2021minimum}.
The system is subject to the following input constraints $\SI{-2}{\meter/\square\second}\leq a_t\leq \SI{2 }{\meter/\square\second}$, $-\frac{\pi}{2} \SI{}{\radian} \leq \delta_t\leq \frac{\pi}{2}  \SI{}{\radian}$
with the initial state $\textbf{x}_0$ and the target state $\textbf{x}_{\text{target}}$ as $[\SI{0}{\meter},\SI{0}{\meter},\SI{0}{\meter/\second},\SI{0}{\radian}]^T$ and $[\SI{201.5}{\meter},\SI{0}{\meter},\SI{0}{\meter/\second},\SI{0}{\radian}]^T$ for each iteration, respectively.
In iteration 0, a brute force algorithm~\cite{rosolia2021minimum} is used to calculate the initial feasible trajectory, which is used by both i2LQR and LMPC algorithms for all simulations.
In iteration 1, both algorithms use the historical data from iteration 0,
and historical data from the two previous iterations are used by both algorithms in subsequent iterations.
In this work, numerical simulation is carried out in Python.
For the LMPC algorithm, CasADi \cite{andersson2019casadi} is used as modeling language and the resulting optimization is solved with IPOPT \cite{biegler2009large}.

\subsection{Iterative Tasks In Static Environments}
\label{Sec:Results-Static}
This subsection compares the performance of our proposed i2LQR and LMPC for iterative tasks in static environments.

In the first group of simulations, no obstacle exists in the environment.
In each iteration, the system travels from the initial state $\textbf{x}_0$ to the target state $\textbf{x}_{\text{target}}$.
The simulation results using the two algorithms are shown in Fig.~\ref{fig:sim-1}.
It indicates that both algorithms minimize the system's completion time to reach the target state and have the same optimal performance.
Specifically, given the same initial trajectory, the trajectories in iteration 10 are almost straight lines between the initial  and target state for both algorithms.
The system accelerates for the first half of the simulation and then decelerate to reach the target state with zero velocity in the second half for both algorithms in iteration 10.
\begin{remark}
    The acceleration profile is not exactly symmetric since the the sampling time $\Delta t$ is set to \SI{1}{\second} which is quite large and cannot be considered as a continuous system anymore. All this could result in multiple optimal trajectories for the same completion time.
\end{remark}

\begin{figure}
    \setlength{\abovecaptionskip}{0.1cm}
    \setlength{\belowcaptionskip}{-0.5cm}
    \centering
    \begin{subfigure}[t]{1.0\linewidth}
        \setlength{\abovecaptionskip}{0cm}
        \setlength{\belowcaptionskip}{0.0cm}
        \centering
        \includegraphics[width=0.94\linewidth]{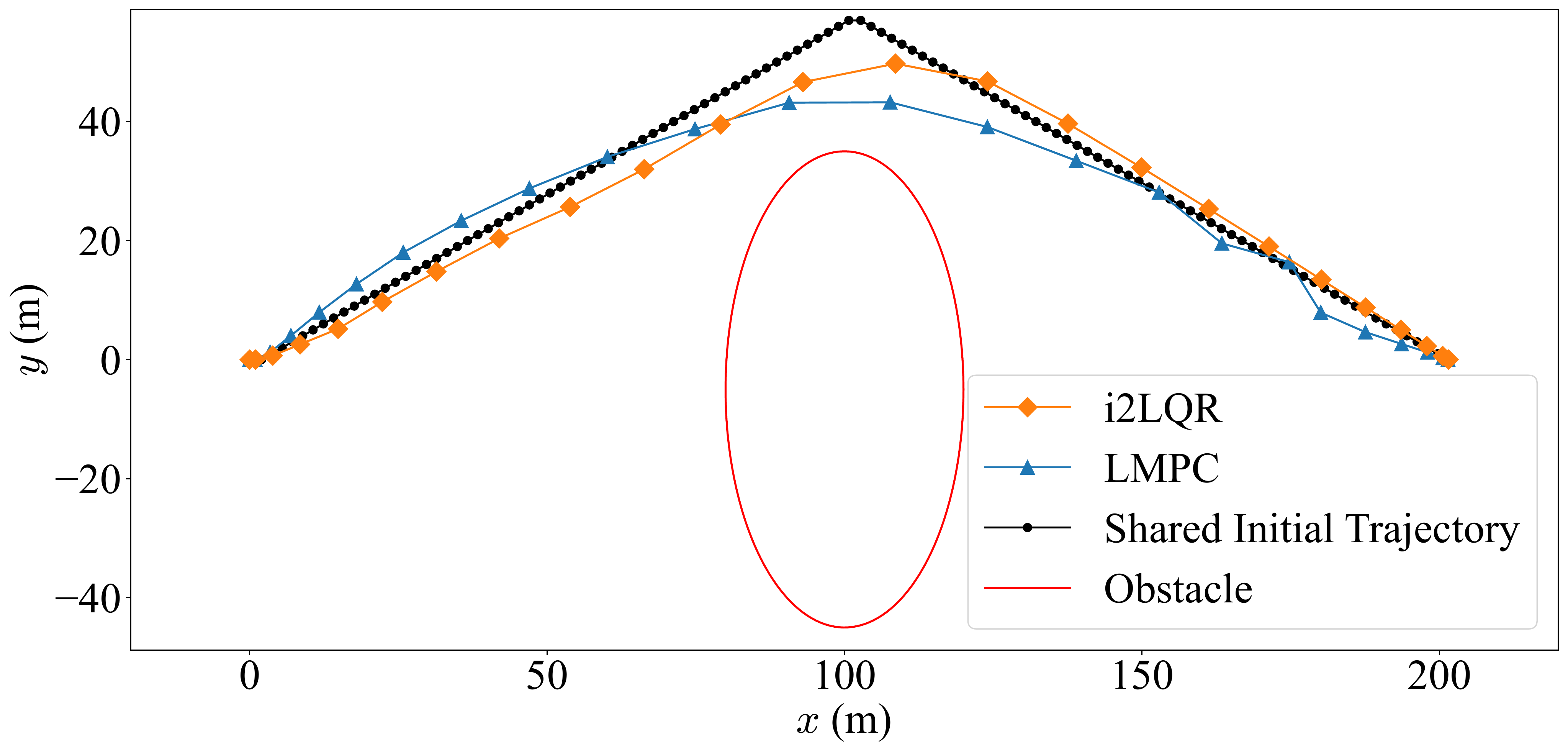}
        \caption{Shared initial trajectory, and trajectories in iteration 10 for i2LQR and LMPC.}
        \label{fig:static_obstacle_trajectory}
    \end{subfigure}
    \begin{subfigure}[t]{1.0\linewidth}
        \setlength{\abovecaptionskip}{0cm}
        \setlength{\belowcaptionskip}{0.0cm}
        \centering
        \includegraphics[width=0.94\linewidth]{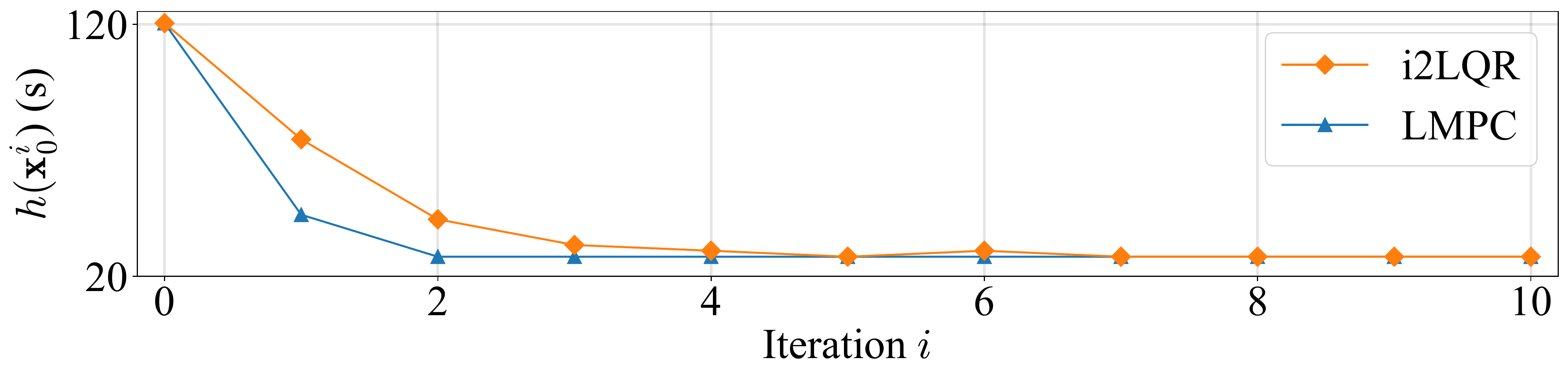}
        \caption{Time to finish the iteration for i2LQR and LMPC.}
        \label{fig:static_obstacle_time}
    \end{subfigure}
    \begin{subfigure}[t]{1.0\linewidth}
        \setlength{\abovecaptionskip}{0cm}
        \setlength{\belowcaptionskip}{0.0cm}
        \centering
        \includegraphics[width=0.94\linewidth]{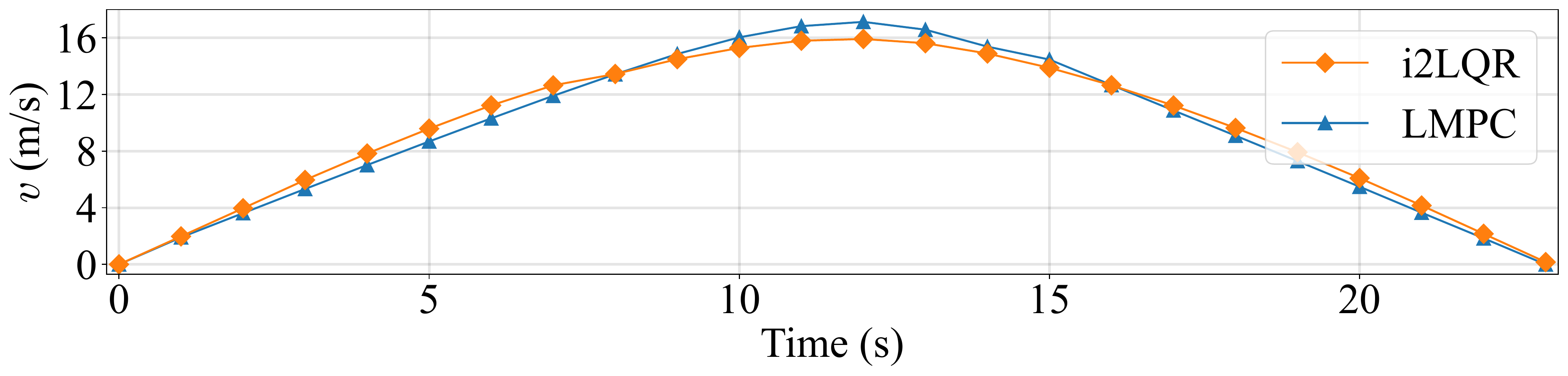}
        \caption{System speed in iteration 10 for i2LQR and LMPC.}
        \label{fig:static_obstacle_spd}
    \end{subfigure}
    \begin{subfigure}[t]{1.0\linewidth}
        \setlength{\abovecaptionskip}{0cm}
        \setlength{\belowcaptionskip}{0.0cm}
        \centering
        \includegraphics[width=0.94\linewidth]{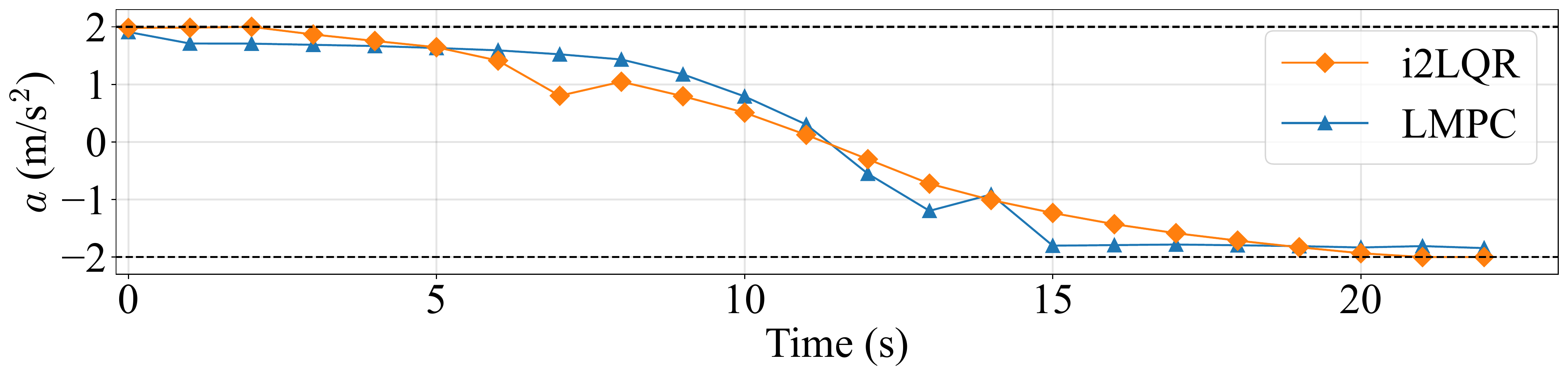}
        \caption{System acceleration in iteration 10 for i2LQR and LMPC.}
        \label{fig:static_obstacle_acc}
    \end{subfigure}
   \caption{Simulation with a static obstacle. Both algorithms could reach the system's optimal performance.}
    \label{fig:sim-2}
\end{figure}

In the second group of simulations, an ellipse-shaped static obstacle with center $(x_{\text{obs}},y_{\text{obs}})=(100\text{ m},-5 \text{ m})$ exists in the environment.
The system must travel from the state $\textbf{x}_0$ to the state $\textbf{x}_{\text{target}}$ while avoiding this obstacle.
Simulation results using the two algorithms are shown in Fig.~\ref{fig:sim-2}.
It's shown that both i2LQR and LMPC algorithms minimize the completion time for the system to reach the target state even when a static obstacle exists in the environment.
Given the same initial trajectory, both algorithms can find the optimal trajectory that avoids the obstacle.
In iteration 10, the system accelerates for the first half of the simulation and then decelerate to reach the target state with zero velocity.

\subsection{Iterative Tasks In Dynamic Environments}
\label{Sec:Results-Dynamic}
To show the proposed i2LQR algorithm's performance for iterative tasks in dynamic environments, we conduct two groups of simulations with different environments. 

\begin{figure}
    \setlength{\abovecaptionskip}{0.1cm}
    \setlength{\belowcaptionskip}{-0.4cm}
    \centering
    \begin{subfigure}[t]{1.0\linewidth}
        \setlength{\abovecaptionskip}{0cm}
        \setlength{\belowcaptionskip}{0.1cm}
        \centering
        \includegraphics[width=0.94\linewidth]{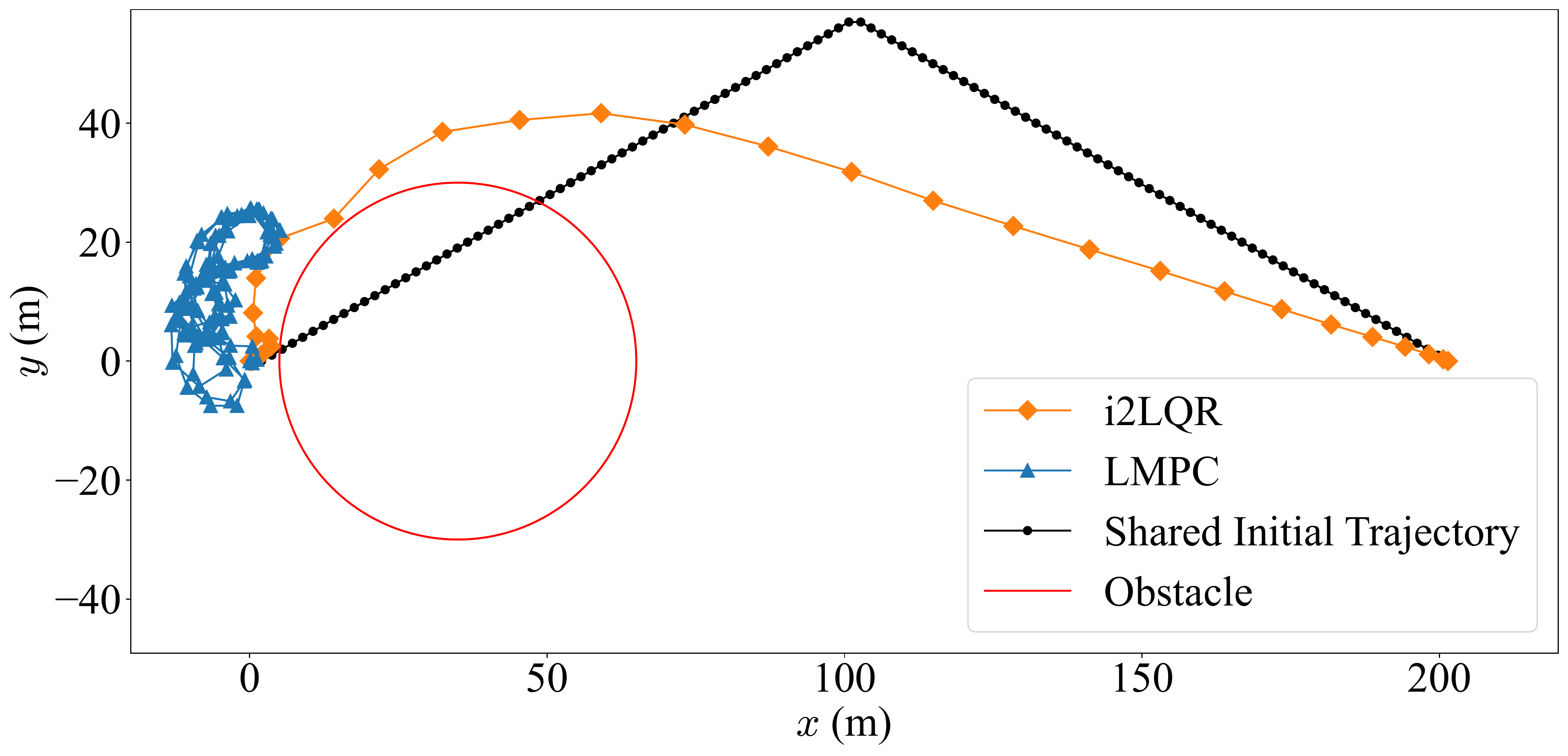}
        \caption{Shared initial trajectory, and trajectories in iteration 6 for i2LQR and LMPC. 
        The static obstacle is plotted in red.}
        \label{fig:add_static_obstacle_trajectory}
    \end{subfigure}
    \begin{subfigure}[t]{1.0\linewidth}
        \setlength{\abovecaptionskip}{0cm}
        \setlength{\belowcaptionskip}{0.1cm}
        \centering
        \includegraphics[width=0.94\linewidth]{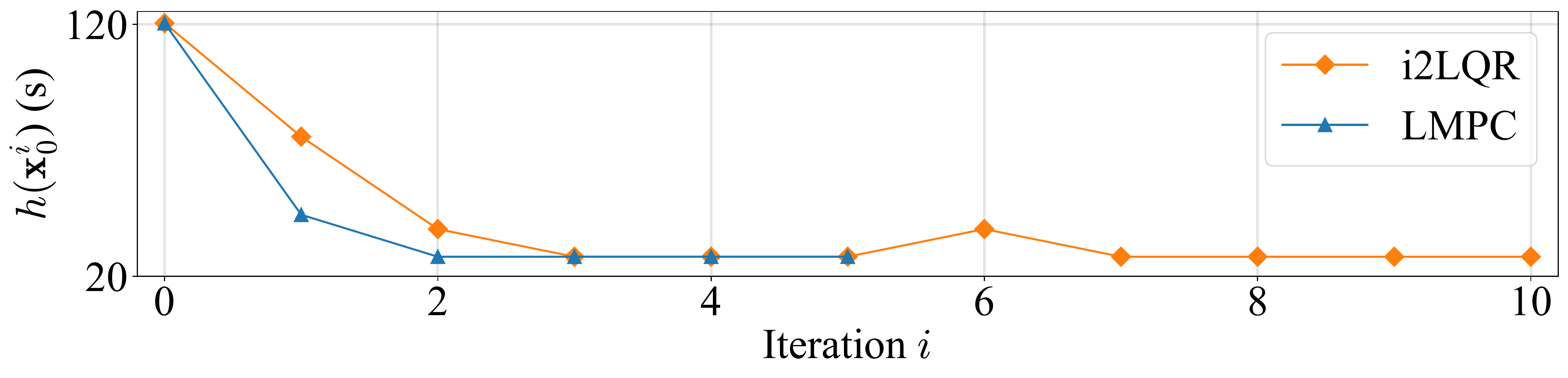}
        \caption{Time to finish the iteration for i2LQR and LMPC.}
        \label{fig:add_static_obstacle_time}
    \end{subfigure}
   \caption{Simulation with an added static obstacle in iteration 6. In iteration 6, the proposed i2LQR reaches the target state $\mathbf{x}_{\text{target}}$ while the LMPC cannot reach the target state $\mathbf{x}_{\text{target}}$.}
    \label{fig:sim-3}
\end{figure}

In the third group of simulations (Fig.~\ref{fig:sim-3}), a static circle-shaped obstacle with center $(x_{\text{obs}},y_{\text{obs}})=(35\text{ m},0 \text{ m})$ exists in the environment for iteration 6.
According to Fig.~\ref{fig:add_static_obstacle_time}, systems with the both algorithms have reached their optimal performance before the static obstacle is introduced.
During iteration 6, i2LQR spends \SI{25}{\second} to reach the target state while avoiding the static obstacle.
Then, it returns to its optimal performance after the obstacle is removed.
However, the LMPC cannot reach the target state after more than \SI{100}{\second}.
The reason is that all the nearby historical states except the initial state are occupied by the obstacle, which results in the infeasibility of the optimization problem. 

\begin{figure}
    \setlength{\abovecaptionskip}{0.3cm}
    \setlength{\belowcaptionskip}{-0.5cm}
    \centering
    \begin{subfigure}[t]{1.0\linewidth}
        \setlength{\abovecaptionskip}{0cm}
        \setlength{\belowcaptionskip}{0.1cm}
        \centering
        \includegraphics[width=0.94\linewidth]{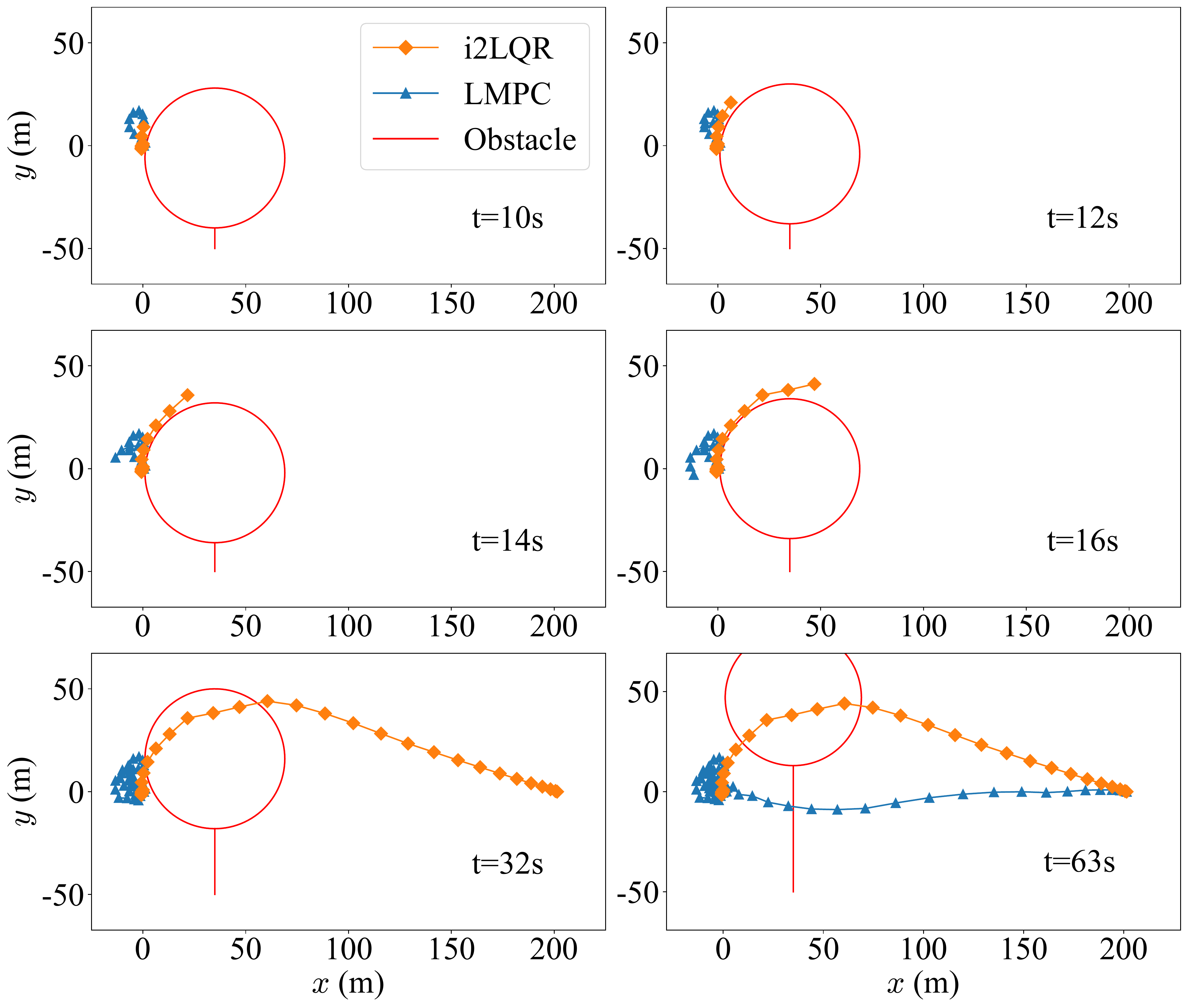}
        \caption{Snapshots for the systems using two algorithms in iteration 6.}
        \label{fig:add_moving_obstacle_trajectory}
    \end{subfigure}
    \begin{subfigure}[t]{1\linewidth}
        \setlength{\abovecaptionskip}{0cm}
        \setlength{\belowcaptionskip}{0.1cm}
        \centering
        \includegraphics[width=0.94\linewidth]{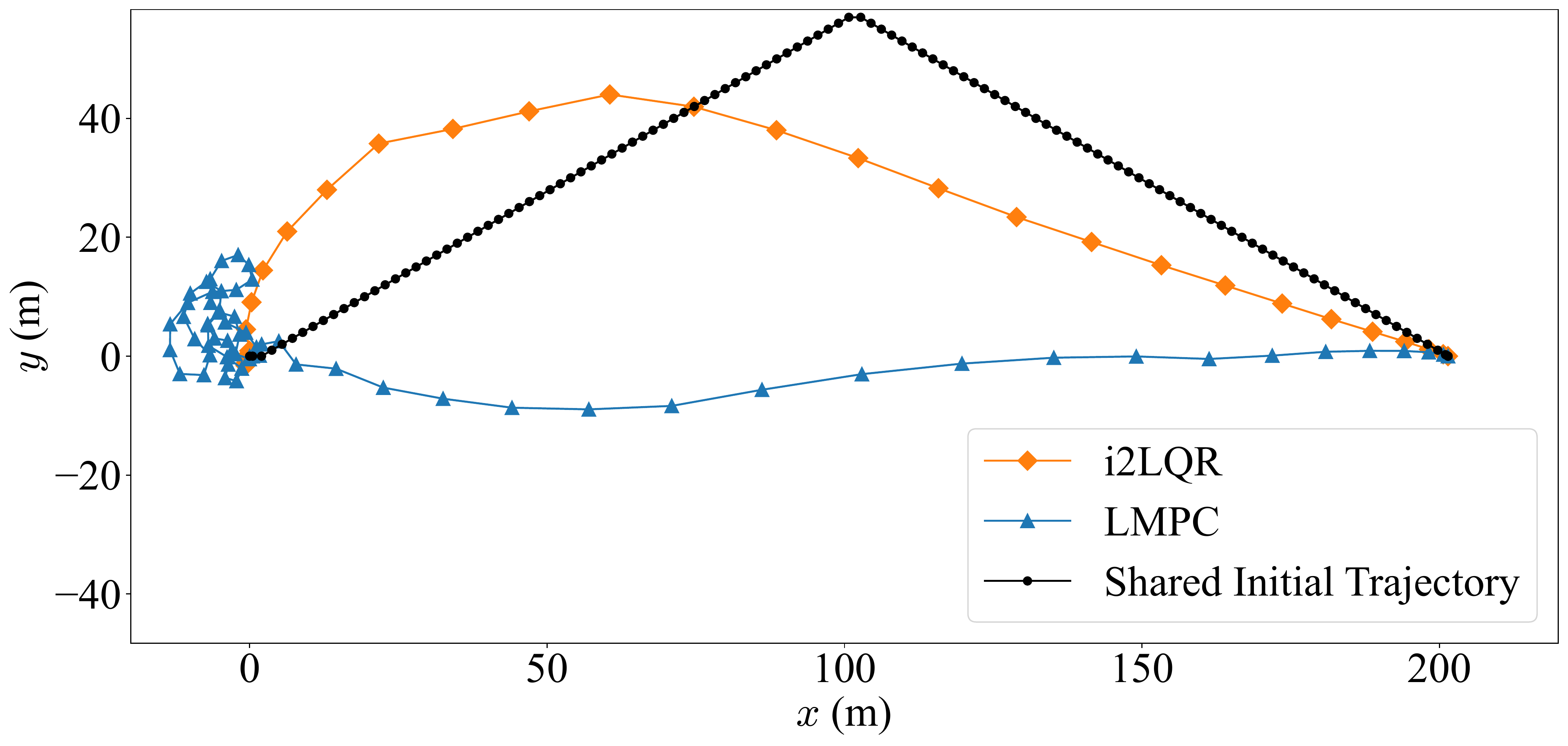}
        \caption{Shared initial trajectory, and trajectories in iteration 6 for i2LQR and LMPC.}
        \label{fig:add_moving_obstacle_total_traj}
    \end{subfigure}
    \begin{subfigure}[t]{1.0\linewidth}
        \setlength{\abovecaptionskip}{0cm}
        \setlength{\belowcaptionskip}{0.0cm}
        \centering
        \includegraphics[width=0.94\linewidth]{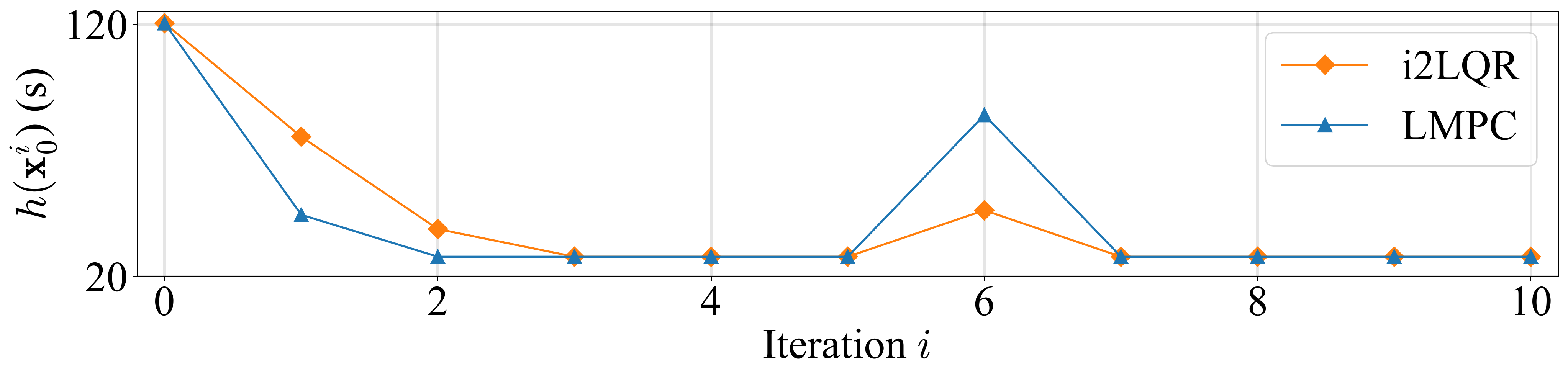}
        \caption{Time to finish the iteration for i2LQR and LMPC.}
        \label{fig:add_moving_obstacle_time}
    \end{subfigure}
   \caption{Simulation with an added moving obstacle in iteration 6. In iteration 6, the proposed i2LQR reaches the target state $\mathbf{x}_{\text{target}}$ earlier than the LMPC, 
   meaning a smaller system's cost-to-go $h(\mathbf{x}^6_0)$.}
    \label{fig:sim-4}
\end{figure}

To further present the proposed i2LQR algorithm's performance in a more complicated dynamic environment, in the fourth simulation, a circle-shaped moving obstacle moves upwards from the initial point $(x_{\text{obs}},y_{\text{obs}})=(35\text{ m},-16 \text{ m})$ with a speed of 1 m/s in iteration 6.
The obstacle is removed in the next iteration.
Fig.~\ref{fig:add_moving_obstacle_trajectory} and Fig.~\ref{fig:add_moving_obstacle_total_traj} show the snapshots and trajectories for both i2LQR and LMPC algorithms in iteration 6, respectively.
It's shown that i2LQR is able to avoid this moving obstacle even when the obstacle is close to the system. 
However, since historical states with smaller time costs are occupied by the obstacle, these states become infeasible for the local MPC optimization of the LMPC algorithm; therefore, the controller cannot drive the system towards the target state at the beginning.
After the obstacle goes away from the system, it moves towards the target state.
The i2LQR spends \SI{32}{\second} to finish iteration 6, while the LMPC needs \SI{63}{\second} to finish this in the same environment.
Both algorithms return to their optimal performance after the moving obstacle is removed from the environment.

\begin{remark}
It's possible to get a solution by adding slack variables to the terminal state constraint of LMPC in \cite{rosolia2021minimum}. 
This converts the hard constraints on the terminal state into cost-based soft constraints. 
However, using slack variables is not in line with the design of the LMPC algorithm, which relies on the feasibility of the terminal state.
Furthermore, this may not guarantee the algorithm's performance.
\end{remark}

%% file: sections/conclusion.tex
\section{Conclusion} 
\label{Sec:Conclusion}
% In this work, a control strategy called Iterative Linear Quadratic Regulator for Iterative Tasks (i2LQR) is presented.
% The proposed algorithm improves the closed-loop performance with local trajectory optimization for iterative tasks in dynamic environments.
% The algorithm utilizes historical data in an optimization problem, and solves it in an iterative manner. 
% To validate our control design, four sets of simulations are conducted. 
% In the first two simulations with static environments, our proposed i2LQR algorithm provides the same optimized performance as state-of-the-art LMPC algorithm. 
% In the remaining two simulations where the environment changes during the simulation, the i2LQR algorithm outperforms the LMPC algorithm.
% In future work, stability and feasibility analysis of the proposed controller will be presented.

This work introduces the Iterative Linear Quadratic Regulator for Iterative Tasks (i2LQR) control strategy, which enhances performance in dynamic environments through local trajectory optimization using historical data. Four simulations are conducted: i2LQR matches state-of-the-art LMPC in static environments and outperforms it in changing environments. Future work will include stability and feasibility analysis of this controller.

%% file: main.bbl
\begin{thebibliography}{10}
\providecommand{\url}[1]{#1}
\csname url@samestyle\endcsname
\providecommand{\newblock}{\relax}
\providecommand{\bibinfo}[2]{#2}
\providecommand{\BIBentrySTDinterwordspacing}{\spaceskip=0pt\relax}
\providecommand{\BIBentryALTinterwordstretchfactor}{4}
\providecommand{\BIBentryALTinterwordspacing}{\spaceskip=\fontdimen2\font plus
\BIBentryALTinterwordstretchfactor\fontdimen3\font minus
  \fontdimen4\font\relax}
\providecommand{\BIBforeignlanguage}[2]{{%
\expandafter\ifx\csname l@#1\endcsname\relax
\typeout{** WARNING: IEEEtran.bst: No hyphenation pattern has been}%
\typeout{** loaded for the language `#1'. Using the pattern for}%
\typeout{** the default language instead.}%
\else
\language=\csname l@#1\endcsname
\fi
#2}}
\providecommand{\BIBdecl}{\relax}
\BIBdecl

\bibitem{jain2020computing}
A.~Jain and M.~Morari, ``Computing the racing line using bayesian
  optimization,'' in \emph{2020 59th IEEE Conference on Decision and Control
  (CDC)}, 2020, pp. 6192--6197.

\bibitem{wu2022model}
X.~Wu, J.~Zeng, A.~Tagliabue, and M.~W. Mueller, ``Model-free online motion
  adaptation for energy-efficient flight of multicopters,'' \emph{IEEE Access},
  vol.~10, pp. 65\,507--65\,519, 2022.

\bibitem{kapania2016sequential}
N.~R. Kapania, J.~Subosits, and J.~Christian~Gerdes, ``A sequential two-step
  algorithm for fast generation of vehicle racing trajectories,'' \emph{Journal
  of Dynamic Systems, Measurement, and Control}, vol. 138, no.~9, 2016.

\bibitem{nagy2019sequential}
A.~Nagy and I.~Vajk, ``Sequential time-optimal path-tracking algorithm for
  robots,'' \emph{IEEE Transactions on Robotics}, vol.~35, no.~5, pp.
  1253--1259, 2019.

\bibitem{heilmeier2019minimum}
A.~Heilmeier, A.~Wischnewski, L.~Hermansdorfer, J.~Betz, M.~Lienkamp, and
  B.~Lohmann, ``Minimum curvature trajectory planning and control for an
  autonomous race car,'' \emph{Vehicle System Dynamics}, 2019.

\bibitem{palleschi2021fast}
A.~Palleschi, M.~Hamad, S.~Abdolshah, M.~Garabini, S.~Haddadin, and
  L.~Pallottino, ``Fast and safe trajectory planning: Solving the cobot
  performance/safety trade-off in human-robot shared environments,'' \emph{IEEE
  Robotics and Automation Letters}, vol.~6, no.~3, pp. 5445--5452, 2021.

\bibitem{gao2020teach}
F.~Gao, L.~Wang, B.~Zhou, X.~Zhou, J.~Pan, and S.~Shen, ``Teach-repeat-replan:
  A complete and robust system for aggressive flight in complex environments,''
  \emph{IEEE Transactions on Robotics}, vol.~36, no.~5, pp. 1526--1545, 2020.

\bibitem{fuchs2021super}
F.~Fuchs, Y.~Song, E.~Kaufmann, D.~Scaramuzza, and P.~D{\"u}rr, ``Super-human
  performance in gran turismo sport using deep reinforcement learning,''
  \emph{IEEE Robotics and Automation Letters}, vol.~6, no.~3, pp. 4257--4264,
  2021.

\bibitem{song2021autonomous}
Y.~Song, M.~Steinweg, E.~Kaufmann, and D.~Scaramuzza, ``Autonomous drone racing
  with deep reinforcement learning,'' in \emph{2021 IEEE/RSJ International
  Conference on Intelligent Robots and Systems (IROS)}, 2021, pp. 1205--1212.

\bibitem{penicka2022learning}
R.~Penicka, Y.~Song, E.~Kaufmann, and D.~Scaramuzza, ``Learning minimum-time
  flight in cluttered environments,'' \emph{IEEE Robotics and Automation
  Letters}, vol.~7, no.~3, pp. 7209--7216, 2022.

\bibitem{kabzan2019learning}
J.~Kabzan, L.~Hewing, A.~Liniger, and M.~N. Zeilinger, ``Learning-based model
  predictive control for autonomous racing,'' \emph{IEEE Robotics and
  Automation Letters}, vol.~4, no.~4, pp. 3363--3370, 2019.

\bibitem{rosolia2017learning}
U.~Rosolia and F.~Borrelli, ``Learning model predictive control for iterative
  tasks. a data-driven control framework,'' \emph{IEEE Transactions on
  Automatic Control}, vol.~63, no.~7, pp. 1883--1896, 2017.

\bibitem{rosolia2021minimum}
------, ``Minimum time learning model predictive control,'' \emph{International
  Journal of Robust and Nonlinear Control}, vol.~31, no.~18, pp. 8830--8854,
  2021.

\bibitem{rosolia2019learning}
------, ``Learning how to autonomously race a car: a predictive control
  approach,'' \emph{IEEE Transactions on Control Systems Technology}, vol.~28,
  no.~6, pp. 2713--2719, 2019.

\bibitem{li2022learning}
G.~Li, A.~Tunchez, and G.~Loianno, ``Learning model predictive control for
  quadrotors,'' in \emph{IEEE International Conference on Robotics and
  Automation}, 2022.

\bibitem{thananjeyan2021abc}
B.~Thananjeyan, A.~Balakrishna, U.~Rosolia, J.~E. Gonzalez, A.~Ames, and
  K.~Goldberg, ``Abc-lmpc: Safe sample-based learning mpc for stochastic
  nonlinear dynamical systems with adjustable boundary conditions,'' in
  \emph{Algorithmic Foundations of Robotics XIV: Proceedings of the Fourteenth
  Workshop on the Algorithmic Foundations of Robotics 14}.\hskip 1em plus 0.5em
  minus 0.4em\relax Springer, 2021, pp. 1--17.

\bibitem{he2022autonomous}
S.~He, J.~Zeng, and K.~Sreenath, ``Autonomous racing with multiple vehicles
  using a parallelized optimization with safety guarantee using control barrier
  functions,'' in \emph{IEEE International Conference on Robotics and
  Automation}, 2022.

\bibitem{chen2017constrained}
J.~Chen, W.~Zhan, and M.~Tomizuka, ``Constrained iterative lqr for on-road
  autonomous driving motion planning,'' in \emph{2017 IEEE 20th International
  Conference on Intelligent Transportation Systems (ITSC)}, 2017, pp. 1--7.

\bibitem{hu2022active}
H.~Hu and J.~F. Fisac, ``Active uncertainty reduction for human-robot
  interaction: An implicit dual control approach,'' in \emph{Algorithmic
  Foundations of Robotics XV: Proceedings of the Fifteenth Workshop on the
  Algorithmic Foundations of Robotics}.\hskip 1em plus 0.5em minus 0.4em\relax
  Springer, 2022, pp. 385--401.

\bibitem{andersson2019casadi}
J.~A. Andersson, J.~Gillis, G.~Horn, J.~B. Rawlings, and M.~Diehl, ``Casadi: a
  software framework for nonlinear optimization and optimal control,''
  \emph{Mathematical Programming Computation}, vol.~11, no.~1, pp. 1--36, 2019.

\bibitem{biegler2009large}
L.~T. Biegler and V.~M. Zavala, ``Large-scale nonlinear programming using
  ipopt: An integrating framework for enterprise-wide dynamic optimization,''
  \emph{Computers \& Chemical Engineering}, vol.~33, no.~3, pp. 575--582, 2009.

\end{thebibliography}
